\documentclass{article}
\usepackage{amsmath}
  \usepackage{paralist}
  \usepackage{graphics} 
  \usepackage{epsfig} 
\usepackage{graphicx}  \usepackage{epstopdf}
\usepackage{subfig}
\usepackage{sidecap}
 \usepackage[colorlinks=true]{hyperref}
	\usepackage[latin1]{inputenc}
\usepackage{tikz}
\usetikzlibrary{shapes,arrows}
	
\hypersetup{urlcolor=blue, citecolor=red}

\begin{document}

\title{A mathematical model about human infections of H7N9 influenza in China with the intervention of live poultry markets closing.}
\author{Xi Huo\thanks{Department of Mathematics, Ryerson University, Toronto, ON M5B 2K3, Canada} \thanks{Center for Disease Modelling, York University, Toronto, ON M3J 1P3, Canada}}

\maketitle
\bigskip

\begin{abstract}
This paper develops a deterministic differential equations model that captures the H7N9 virus transmission from live poultry to human via poultry-human contacts in live poultry markets (LPMs). The virus circulation among live poultry, which happens but is hard to be detected (since contaminated live poultry appear to be asymptomatic), is also incorporated in the model. The time-dependent contact rate between human and live poultry based on LPMs closing information can be estimated. From data of LPMs closing news, the contact rate function can be easily estimated. This model could serve as a rational basis for public health authorities to evaluate the effectiveness of LPM closing, as well as other interventions according to simple modifications. Without data about daily cases, I also provide suggestions for some of the basic parameters that would be a useful fitness parameter set for future simulation.
\end{abstract}

\begin{section}{Background and Methods}

The dynamic of the H7N9 virus transmission is similar to that of a nosocomial bacteria infection \cite{webb}, from which was the modeling idea originated. Figure \ref{compartment} shows the dynamics of H7N9 transmission from live poultry to human and among live poultry. We divide human population into three classes: susceptible people who frequently go to LPMs, H7N9 infected people, and people identified of H7N9 infection. Since the total number of live poultry is hard to be estimated, we use fraction to measure contaminated and uncontaminated live poultry. For simplicity, we assume homogeneous mixing among live poultry, as well as homogeneous contact between susceptible people and live poultry in the LPMs.\\

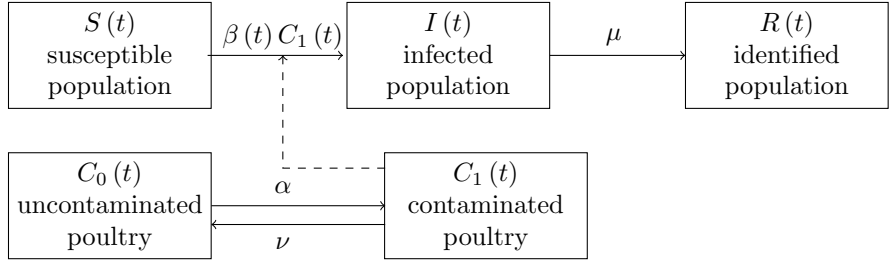
\begin{figure}[ht]
\tikzstyle{block} = [rectangle, draw, fill=white, 
    text width=7em, text centered, minimum height=3em]
\begin{tikzpicture}
    \node (S) at (0,0) [block] {$S\left(t\right)$\\susceptible population};
		\node (beta) at (2.3,0.25) {$\beta\left(t\right) {C_1}\left(t\right)$};
    \node (I) at (4.5,0) [block] {$I\left(t\right)$\\infected population};
		\node (mu) at (6.7,0.25) {$\mu$};
    \node (R) at (9,0) [block] {$R\left(t\right)$\\identified population};
    \node (C0) at (0,-2) [block] {${C_0}\left(t\right)$\\uncontaminated poultry};
		\node (alpha) at (2.3, -1.75) {$\alpha$};
    \node (C1) at (5,-2) [block] {${C_1}\left(t\right)$\\contaminated poultry};
		\node (nu) at (2.3, -2.5) {$\nu$};
    \draw [->] (1.3,0) -- (3.1,0);
		\draw [->] (5.85,0) -- (7.65,0);
    \draw [->] (1.35,-2) -- (3.66,-2);
    \draw [->] (3.66,-2.25) -- (1.35,-2.25);
		\draw [->, dashed] (3.66,-1.5) -| (2.3,0);
\end{tikzpicture}
\caption{\label{compartment} Compartmental model of transmission dynamics. $S(t)$ is the number of susceptible individuals at time $t$, $I(t)$ is the number of infected people at time $t$, $R(t)$ is the number of people who become symptomatic at time $t$. ${C_0}(t)$ is the fraction of uncontaminated live poultry at time $t$, ${C_1}(t)$ is the fraction of contaminated live poultry at time $t$. Hence, ${C_0}(t)+{C_1}(t)=1$. $\beta\left(t\right)$ is the transmission rate of H7N9 virus from live poultry to human, which is a product of poultry-human contact rate and the possibility for a transmission to occur during contact.}
\end{figure}

Based on the observation of H7N9 population outbreak, the circumstance of constant poultry-poultry transmission rate can be ruled out. The ordinary differential equations system of the model \eqref{ODE_constant} is presented in Appendix. With a constant poultry-poultry transmission rate $\alpha$, the basic reproduction number of the live poultry system is ${\Re _0} = \frac{\alpha }{\nu }$. Depending on initial conditions, one of the following dynamics must have happened: (1) if ${\Re _0} \leq 1$, the fraction of contaminated live poultry decreases to zero and stabilize after sufficiently long time; if ${\Re _0} > 1$, the fraction (2) either decreases to a non-zero value and stabilize after sufficiently long time, or (3) increases to a non-zero value and stabilize after sufficiently long time. With a decreasing poultry-human transmission rate function \cite{LPMinfo} estimated for the outbreak in spring, 2013, the above three possibilities would lead to either of the following happens: (1) and (2) would both give a decreasing number of daily symptomatic people through out the outbreak, which does not match the outbreak result reported by WHO \cite{WHO}; (3) would possibly match the outbreak \cite{WHO} until the resume of LPMs, but could not explain the fact of no new cases appeared after the resume of LPMs in late May, 2013. Therefore, we assume that the poultry-poultry contamination rate, $\alpha$, to be a time-dependent parameter, then the reduced system follows:

\begin{flalign}\label{ODE}
\begin{split}
& {{dS\left( t \right)} \over {dt}} = - \beta\left(t\right) {C_1}\left( t \right)S\left( t \right)\\
& {{dI\left( t \right)} \over {dt}} = \beta\left(t\right) {C_1}\left( t \right)S\left( t \right) - \mu I\left( t \right)\\
& {{d{C_1}\left( t \right)} \over {dt}} = {C_1}\left( t \right)\left[ {\left( {\alpha \left( t \right) - \nu } \right) - \alpha \left( t \right){C_1}\left( t \right)} \right]
\end{split}
\end{flalign}

With the time-dependent contamination rate $\alpha \left( t \right)$, if $\frac{{\alpha \left( t \right)}}{\nu } < 1$ after a certain time, the epidemic will finally stop; otherwise, if $\frac{{\alpha \left( t \right)}}{\nu } > 1$ after a certain time, new cases would consistently appear. $\alpha \left( t\right)$ would fluctuate due to impacts such as seasonality and some unknown aspects, hence in the simulation, people should focus more on the system behavior in relatively short terms that are no longer than $6$ months.
\end{section}

\begin{section}{Simulation guidelines}

Provided with data about the two outbreaks, say 2013 spring and 2013-2014 winter, the poultry contamination rate function $\alpha \left ( t\right)$ can be backward-calculated from the parameters estimated above. Then we can use our results about $\alpha \left( t \right)$ to predict the outbreak in 2013-2014 winter, in order to testify the model. Here as an example, two outbreaks in two specific areas in China, 2013 spring outbreak in Yangtze Delta River Region and 2013-2014 winter in Guangdong Province, are presented for the ways to estimate parameters.\\

Table \ref{parameters} summarizes the definitions and estimated values of the parameters in our model for each of the two simulations: Yangtze Delta River Region, spring 2013 and Guangdong Province, winter 2013 - spring 2014. \cite{humanbehavior} provides percentages of people that frequently visited LPMs in spring 2013 in several main cities of China. Susceptible population for each scenario is estimated by multiplying the total population \cite{yearbook} and the corresponding percentage of the main city in the region. The baseline value of transmission rate function $\beta$ is estimated as the attack ratio, then the $\beta$ function values at each stage of LPM-closing are prorated from the baseline value according to the population of the LPM-closing area. A backward calculation based on some outbreak data could be performed, to help estimate the poultry-poultry contamination rate function $\alpha \left(t\right)$. And we can then use the same estimation to simulate the outbreak in Guangdong, winter 2013 - spring 2014.
\begin{table}[ht]
\begin{center}
\begin{tabular}{|p{5cm} p{1.5cm} p{5cm}|}
\multicolumn{3}{l}{}\\
\hline
Parameter & Value & Sources\\
\hline
removal rate of infectives $\mu$ & to be determined
 & average length of incubation period\\
\hline
birth Rate of Poultry $\nu$ & ${{{{40}^{ - 1}}} \mathord{\left/
 {\vphantom {{{{17}^{ - 1}}} {day}}} \right.
 \kern-\nulldelimiterspace} {day}}$ & average lifespan of farm-raised chicken \cite{roleofpoultry}\\
\hline
 \multicolumn{3}{|c|}{Yangtze River Delta Region, 2013}\\
\hline
initial susceptible population ${S_0}$ & $6 \times {10^7}$ & \cite{originalinfectionsource}, \cite{humanbehavior}\\
transmission rate function $\beta$ & Figure \ref{beta1}& \cite{LPMinfo}, \cite{yearbook}\\
\hline
 \multicolumn{3}{|c|}{Guangdong Province, Winter 2013 - Spring 2014}\\
\hline
initial susceptible population ${S_0}$ & $5 \times {10^7}$ & \cite{yearbook}, \cite{humanbehavior}\\
transmission rate function $\beta$ & Figure \ref{beta2} & \cite{caaa}, \cite{yearbook}\\
\hline
\end{tabular}
\end{center}
\caption{\label{parameters} Parameter Values.}
\end{table}

\begin{figure}
\subfloat[Poultry-human transmission rate function $\beta\left(t\right)$ in Yangtze Delta River Region, spring 2013. \label{beta1}]{
\includegraphics[scale=0.45]{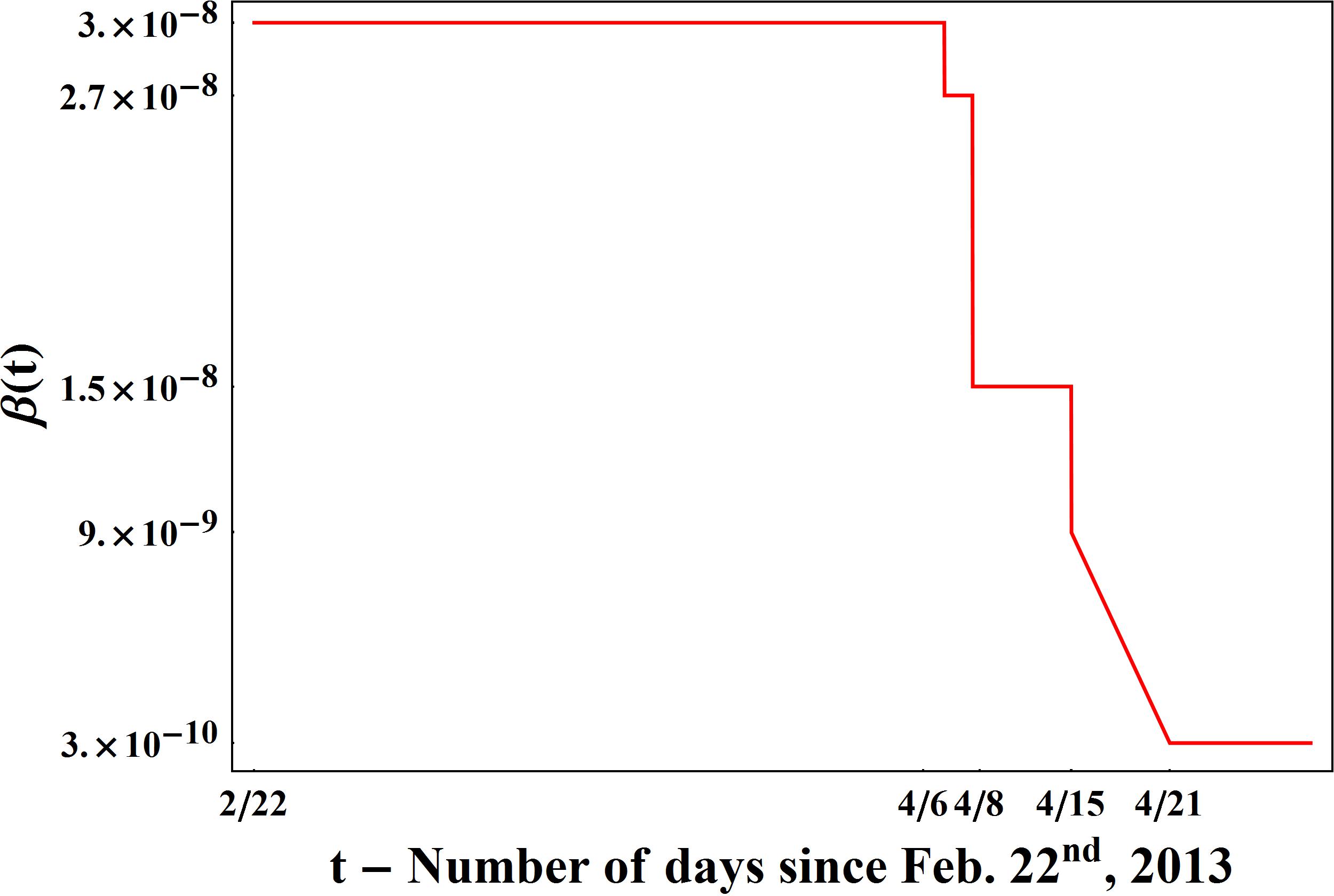}
}
\hfill
\subfloat[Poultry-human transmission rate function $\beta\left(t\right)$ in Guangdong Province, winter 2013- spring 2014. \label{beta2}]{
\includegraphics[scale=0.48]{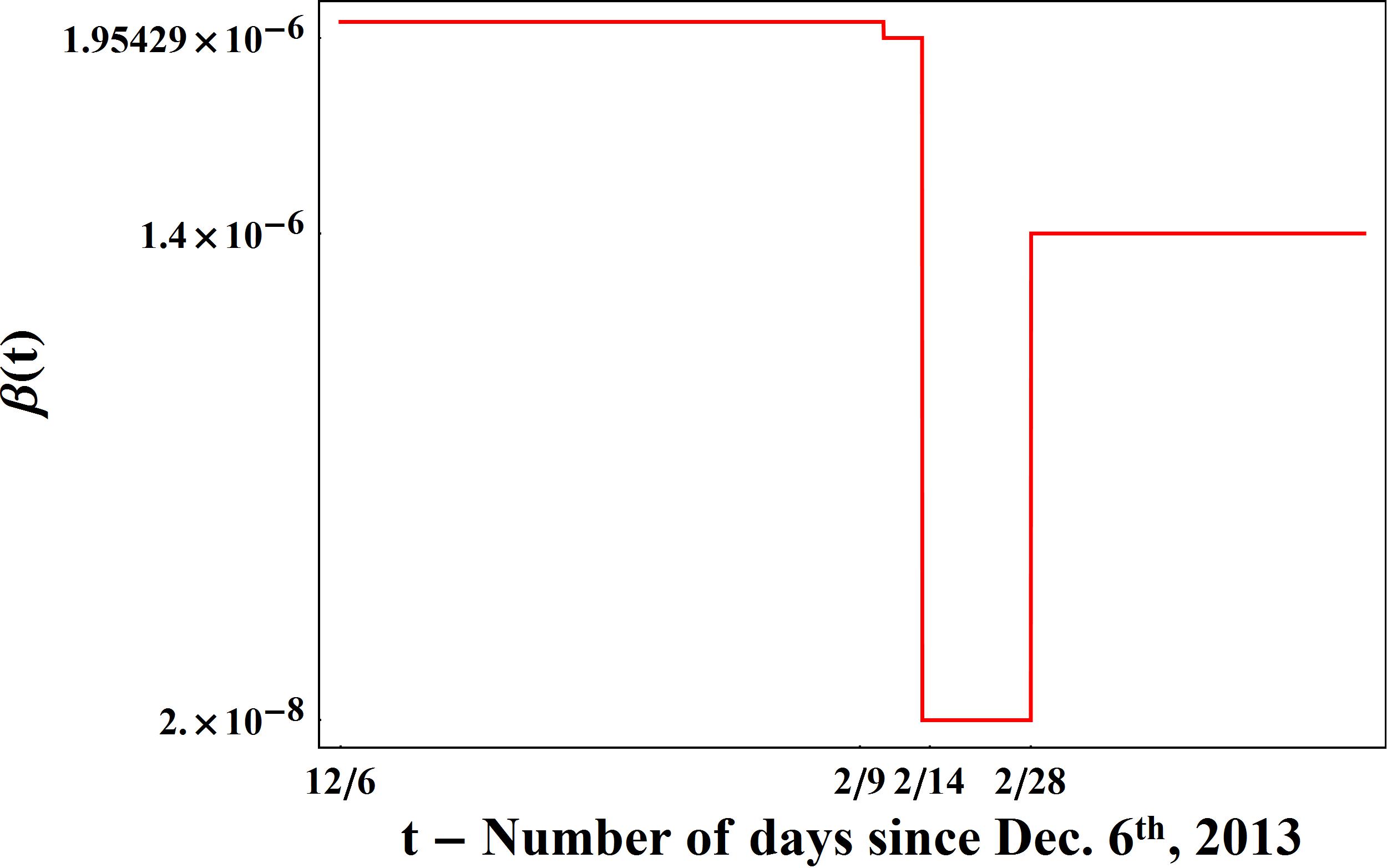}
}
\caption{Figure \ref{beta1}: key dates are marked on the time axis: Feb. 22nd is the beginning date of the simulation, Apr. 6th , Apr. 8th, Apr. 15th , and Apr. 21st are the dates of closing of LPMs in different areas, and the function value is prorated in terms of the population in each area. Figure \ref{beta2}: key dates are marked on the time axis: Dec. 6th is the beginning date of the simulation, Feb. 10th is the first date of LPM closing in Zhongshan, Feb. 15th is the first date of the 2-week LPM closing for the whole province, Mar. 3rd is the first date of LPM resuming and the function value is reduced by $30\%$ after the resuming as reported \cite{caaa}.
}
\end{figure}

As another way to testify our simulations is, we could compute the dynamics of contaminated live poultry ratio in the two outbreaks, especially in 2013-2014 winter, and compare the ratio with that announced officially on Mar. 11th, 2014, by a random detection by Chinese government public health authorities. The match of the data could be another aspect of the reality to tell us if the model and the parameters make sense.\\
\end{section}

\begin{section}{Discussion}
The model used in this research can be easily extended to consider the case of human to human transmission when it occurs, then more public health interventions should be taken into consideration, such as isolation, quarantine, and contact tracing. Our model could be a start in the future if the virus evolves.\\

People's fears of avian flu not only refer to the recently outbreak in some regions and the wide spread, what appears to be more dangerous is the possible emergence of the virus and its future ability to be transmitted from human to human (the current outbreak of Ebola virus in African countries is the ultimate nightmare of the viral emergence). Preventing the spread from animals to human is one possible way to avoid such fears from happening. How efficient the intervention of cutting the animal source of spreading the disease, such as closing LPMs in H7N9 case, can be testified given real data by the model presented.
\end{section}

\begin{section}{Appendix}
\begin{flalign}\label{ODE_constant}
\begin{split}
& {{dS\left( t \right)} \over {dt}} = - \beta\left(t\right) {C_1}\left( t \right)S\left( t \right)\\
& {{dI\left( t \right)} \over {dt}} = \beta\left(t\right) {C_1}\left( t \right)S\left( t \right) - \mu I\left( t \right)\\
& {{dR\left( t \right)} \over {dt}} = \mu I\left( t \right)\\
& {{d{C_0}\left( t \right)} \over {dt}} =  - \alpha {C_0}\left( t \right){C_1}\left( t \right) + \nu {C_1}\left( t \right)\\
& {{d{C_1}\left( t \right)} \over {dt}} = \alpha {C_0}\left( t \right){C_1}\left( t \right) - \nu {C_1}\left( t \right)
\end{split}
\end{flalign}
\end{section}

\end{document}